\documentstyle[11pt,newpasp,twoside,epsf]{article}
\def\etal{{et al.\ }}

\def\hmpc{${h^{-1}}${\rm Mpc}}
\def\hMsun{{h^{-1}}{\rm M}_{\solar}}
\def\solar{\ifmmode_{\mathord\odot}\;\else$_{\mathord\odot}\;$\fi}
\def\LCDM{$\Lambda$CDM}

\begin{document}
\bibliographystyle{abbrv}
\title{Clustering of high-redshift galaxies: relating LBGs to dark matter halos}

\author{Risa H. Wechsler\altaffilmark{1}, 
	James S. Bullock\altaffilmark{1}, 
	Rachel S. Somerville\altaffilmark{2}, 
	Tsafrir S. Kolatt\altaffilmark{1,2}, 
	Joel R. Primack\altaffilmark{1}, 
	George R. Blumenthal\altaffilmark{3}, 
	and Avishai Dekel\altaffilmark{2}}

\altaffiltext{1}{Physics Department, University of California, Santa Cruz, 
	CA 95064}
\altaffiltext{2}{Racah Institute of Physics, The Hebrew University, 
	Jerusalem 91904 Israel}
\altaffiltext{3}{Astronomy \& Astrophysics Department, University of 
	California, Santa Cruz, CA 95064}

\begin{abstract}

We investigate the clustering properties of high-redshift galaxies
within three competing scenarios for assigning luminous galaxies to
dark matter halos from N-body simulations: a one galaxy per massive
halo model, a quiescent star formation model, and a collisional
starburst model.  We compare these models to observations of
Lyman-Break galaxies at $z\sim 3$.  With current data and the simple
statistic used here, one cannot rule out any of these models, but we
see potential for finding distinguishing features using statistics
that are sensitive to the tails of the distribution, and statistics
based on the number of multiple galaxies per halo, which we explore in
an ongoing study.

\end{abstract}


\section{Introduction}

Recent years have seen a new wealth of data on high-redshift galaxies.
In particular, Steidel and collaborators have pioneered an efficient
technique for identifying a population of galaxies between redshifts
of $z \sim 2.5-3.5$, and more recently out to $z\sim 4.5$.  These
galaxies are known as the Lyman-break galaxies (LBGs) due to their
color-selection at the Lyman-break.

One of the first things noticed about these galaxies was that they are
highly clustered, and several workers have investigated whether the
clustering properties of these galaxies could be used to distinguish
between various cosmologies.  In Wechsler \etal (1998, hereafter W98)
we compared the clustering of LBGs observed by Steidel \etal (1998,
hereafter S98) to N-body simulations of 4 different CDM
cosmologies. W98 and a number of other authors (e.g., Katz \etal 1998,
Adelberger \etal 1998, hereafter A98) have found that the clustering
properties of Lyman-break galaxies can be well matched by most
reasonable cosmologies.

Although it may be difficult to distinguish between cosmologies with
this high-redshift data, there is still the hope that these galaxies
may have much to teach us about galaxy formation.  Most of the
previous investigations of this subject have made simple assumptions
about the relation between dark matter halos and visible galaxies, and
found that the data can be well-matched by assuming a one-to-one
correspondence between galaxies and the most massive halos.  However,
the detailed relationship is probably not this simple; in this
proceeding we investigate the effects of relaxing this assumption, and
focus on the question of how the clustering properties of the LBGs can
help us to understand their relation to dark matter halos and
constrain models of galaxy formation.  In this proceeding, we use
N-body simulations to calculate the expected clustering properties of
high-redshift galaxies under different assumptions for their halo
occupation, motivated by several popular models of galaxy formation.
Before doing this, however, it is useful to investigate a simple model
in order to gain understanding of how varying assumptions can affect
the LBG clustering properties.

There have been two general scenarios proposed for the nature of the
LBGs.  In one, known as 'Central Quiescent' (see, e.g., S98, A98, 
Governato \etal 1998, Baugh \etal 1998)
the LBGs form only at the centers of the most massive galaxies, which
slowly accrete gas and form stars.  In the other, known as
'Collisional Starburst' (e.g., Lowenthal \etal 1997, Somerville, Primack \&
Faber 1999, hereafter SPF, Kolatt \etal 1999)
most of the LBGs are formed by small
colliding galaxies that trigger a burst of star formation.
Consider two simple models for the halo occupation in the above
scenarios: first, for central quiescent LBGs, allow one LBG per halo
above a mass threshold $M_{min}^{CQ}$, and second, for collisional
starburst LBGs, let the average number of LBGs per halo of mass
$M>M_{min}^{CS}$, scale like 
$n(M) \propto M^S$, where $S=1.3$ --- this choice is motivated further
in \S 4.  In the first case, since LBGs are found only in the most
massive halos, their clustering will be biased with respect to the
underlying dark matter distribution.  In the second case, although
collisions can be found in small-mass halos (typically $M_{min}^{CS} <
M_{min}^{CQ}$), they will preferentially be located within large
halos, and the distribution of collisions should be also strongly
biased.

A quantitative estimate of this bias can be obtained using the model
of Mo \& White (1996), who give an expression for the halo bias
$b_{\rm h} = \xi_{\rm h}/\xi_{\rm DM}$.  The appropriate bias factor
for a population of galaxies within halos more massive than $M_{min}$
is an average of $b_h(M,z)$ weighted by the abundance of halos as a
function of mass, $dN_h/dM$ (e.g., as estimated by the Press-Schechter 1974
approximation), and the average number of galaxies per halo $n(M)$:
\begin{equation}
b_g(z, M>M_{min}) = \frac{1}{N_g(z)} \int_{M_{min}}^{\infty}
\frac{dN_h}{dM}(M,z) b_h(M,z) n(M) dM,
\label{eqt:bg}
\end{equation}
where $N_g(z) = \int_{M_{min}}^{\infty} \frac{dN}{dM}(M,z) n(M) dM$.
For the simple central quiescent scenarios discussed above, $n(M)=1$
and we obtain the standard expression for halo bias.  We use
expressions for $dN_h/dM$ and $b_h$ supplied by Sheth \& Tormen
(1999).

In Figure 1a, we plot $b_g(M_{min})$ at $z=3$ for both scenarios
assuming the \LCDM\ cosmology discussed below.  Because high-mass
halos are weighted more strongly in the collisional starburst model,
we expect galaxies to be more biased for fixed $M_{min}$.  If we are
free to set the mass threshold, we see that typical collisional
starburst scenarios ($M_{min}^{CS} \sim 10^{11} \hMsun$) and central
quiescent scenarios ($M_{min}^{CQ} \sim 10^{12} \hMsun$) can yield
similar clustering properties.  This is shown explicitly using N-body
simulations by Kolatt et al. (1999).

\begin{figure}[htb]
\plottwo{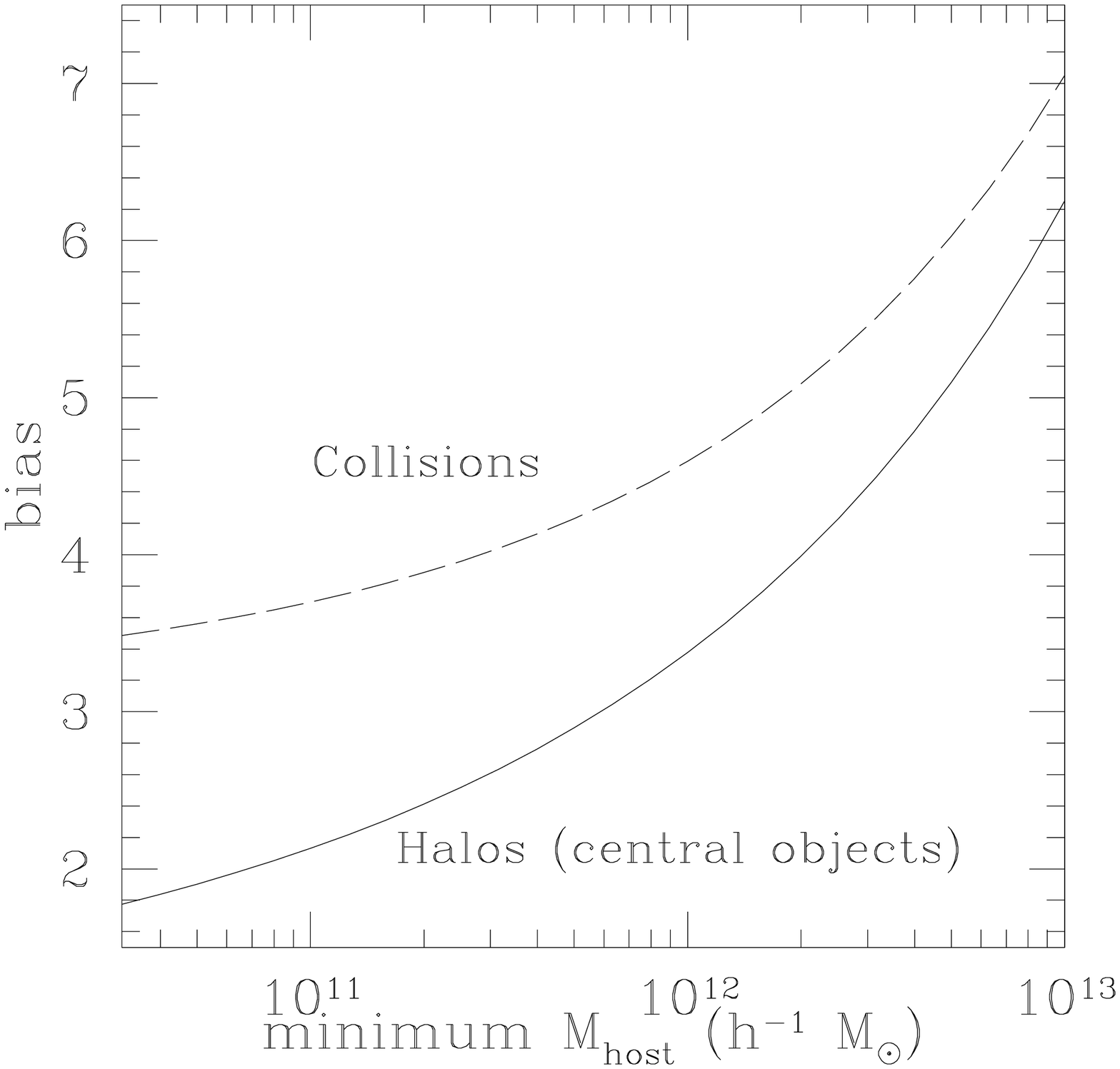}{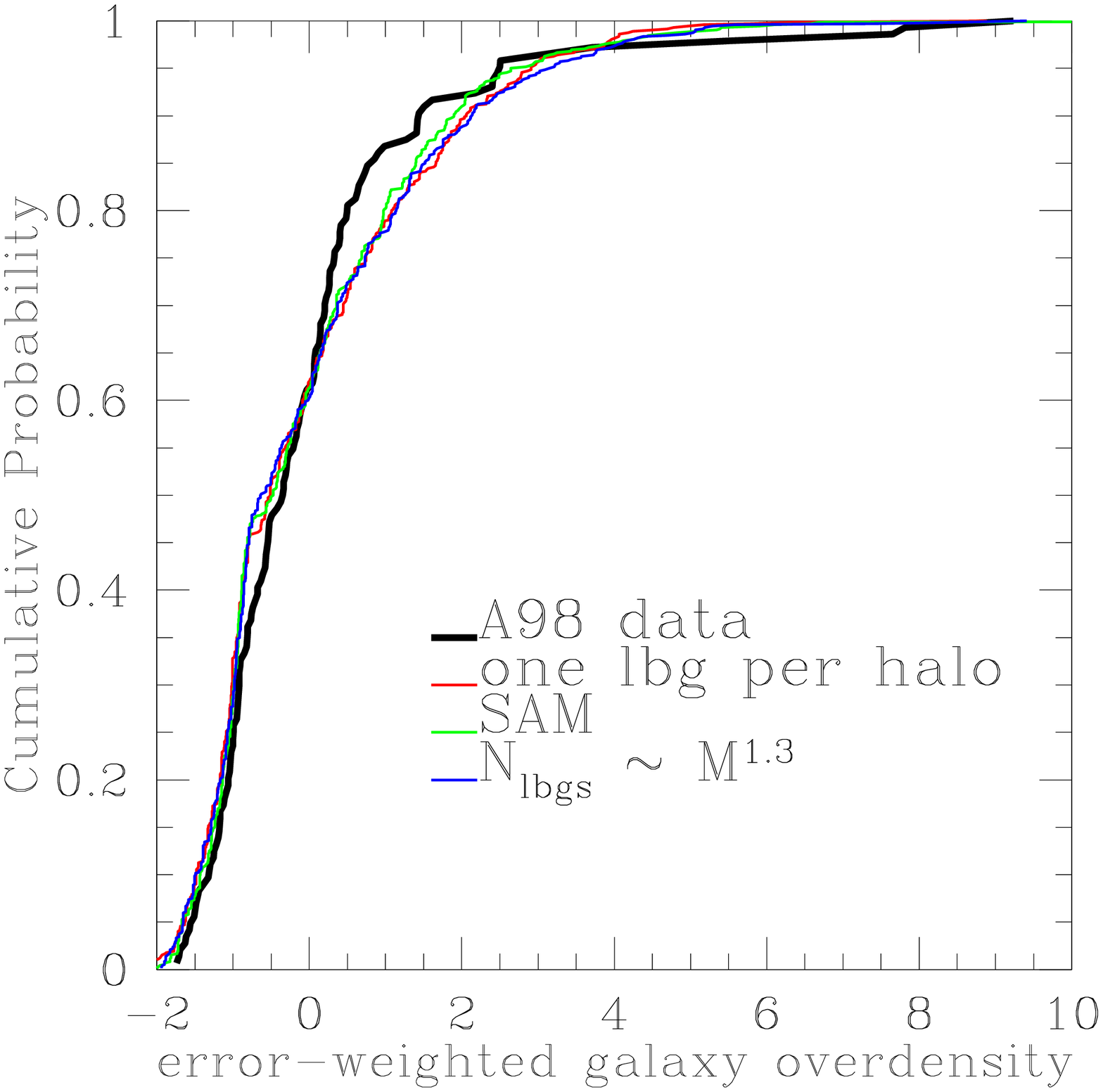}
\caption{Left: The bias parameter for halos and collisions as a function
of $M_{min}$.  Right: The cumulative distribution of the error-weighted galaxy
overdensity, for three LBG-halo models 
described in \S4, compared with the data of A98.}
\label{fig:b}
\end{figure}

\section{Data}
In the present analysis, we compare to the most recent published data
of A98, who have found 376 galaxies in six $9\arcmin \times 9 \arcmin$
fields with redshifts between 2.5 and 3.5.  In analyzing this data,
and 'observing' the simulations,
we use a selection function fit to a histogram of all data, assume that 100\%
of galaxies with $R >25.5$ are photometric candidates at $z \sim 3$, 
and assume that $\sim 45\%$ of candidates have spectroscopic redshifts.

\section{Simulations}

The results we present here are based on N-body simulations of the GIF
collaboration, which were carried out at the Max-Planck-Institut
f\"{u}r Astrophysik, Garching and the Edinburgh Parallel Computing
Centre using codes kindly made available by the Virgo Supercomputing
Consortium.  The halo catalogues have been used for other purposes in,
e.g., Kauffmann \etal 1999.  Here we focus on one cosmology, an \LCDM\
model with $\Omega_m = 0.3$ and $h = 0.7$.  We use dark matter halos
at $z= 3$ from a simulated box which is 141 \hmpc\ on a side.  We then
test a series of models for populating these dark matter halos with
visible galaxies, detailed in the following section.

Once the halos have been populated with galaxies, we observe these galaxies
exactly as the data is observed, making the same assumptions about the
selection function mentioned above.  We normalize each model to get the 
correct number density of observed galaxies.  This normalization is different
for each type of model, and is discussed further below.

\section{Models for populating halos}

{\bf (1) One LBG per massive halo:}
This simple model for populating halos with Lyman-Break galaxies is
motivated by the assumption that the luminosity of these galaxies is
directly proportional to mass, and has been considered by several
authors (e.g., W98, Jing \& Suto 1998).  In this model the number density is
fixed to match the data by choosing an appropriate $M_{min}$.

{\bf (2) Semi-analytic models with quiescent star formation:}
Since clearly the model discussed above is somewhat simplistic, we
consider a more accurate way to fill halos with quiescently
star-forming galaxies, using the semi-analytic models of SPF, which
give the probability of observing galaxies of various luminosities
given the mass of the halo.  Here we only consider one model, referred
to as SFR-M in SPF; see Wechsler \etal (1999, hereafter W99) for
consideration of a wider range of models.  In this class of models,
the number density of objects is normalized using an adjustable dust
parameter described in SPF \& W99.
The dust parameter we use is slightly greater
than that estimated by Steidel \etal (1999) for these galaxies.

{\bf (3) A toy model for collision-driven starbursts:}
In the collisional starburst picture, most of the
visible high-redshift galaxies are starbursting due to recent
collisions.  We have analyzed a very high-resolution (ART) 
N-body simulation of the same \LCDM\ model
within a 30 $h^{-1}$Mpc  box (Kravstov, Klypin \& Khokhlov 1997)
where halo substructure and mergers have been identified
(for details see Bullock \etal 1999, Kolatt \etal 1999).
Using collisions identified in
this simulation, we have found that the number of collisions per host
halo, $n_{\rm coll}$, follows a power law with host halo mass $M_{\rm
host}$,
$n_{\rm coll} \propto M_{\rm host}^S,$
with $S \simeq 1.3$.  For further details on how this model is fit
from simulations and how it might be explained, see W99. 
We use this power-law model to populate
the GIF dark matter halos with collisional starburst galaxies.
We take the minimum mass for a halo
to host a collision to be $M_{min} = 10^{11} h^{-1}M_{\odot}$; the number
density for this model is set by adjusting the constant in front of
the power law.

\section{Comparison of models with data}

Here we just consider one statistic for comparing our models with the
current data; for a more thorough analysis see W99.
A standard statistic for measuring the clustering of a population is
the overdensity in some region; here we consider the overdensity in
cells that are $\delta z = 0.04$ in redshift and $9\arcmin \times 9\arcmin$ in
angle.  In order to minimize the contribution of cells where the
selection function is very small, we instead look an error-weighted
overdensity per pixel, defined as:
\begin{equation}
D_i =\delta N_i/\sigma_i = \delta N_i/N_i^{\onehalf},
\end{equation}
where $N_i$ is the number per pixel and  $\delta N_i$ = ($N_i-\bar{N})/\bar{N}$.

For each model, we compare the cumulative distribution of this
statistic with that of the data (Figure 1b), and there is some
indication that the collisional model has an excess of both overdense
and underdense cells relative to the data.  A detailed
Kolmogorov-Smirnov test for each model, which gives the probabilities
that the data and the model came from the same underlying
distribution, yields:\\
\indent model (1): $P_{KS} =  0.15 \pm 0.06$\\
\indent model (2): $P_{KS} =  0.23 \pm 0.10$\\
\indent model (3): $P_{KS} =  0.10 \pm 0.04$\\
In this analysis, there are 144 data pixels and 720 simulation pixels;
we do the assignment of galaxies to halos and 'observation' of LBGs 10
times for each model to get an error estimate.  

\section{Discussion}

The results above show that none of the three models can be rejected
with this statistic and the current data, but there are indications
that the different models may provide distinguishable 
clustering properties.  In W99, we consider both a
wider range of models and more discriminating statistics, 
including the number of close pairs and a Kuiper test, 
sensitive to distribution tails.

\acknowledgments
{We thank Kurt Adelberger, Sandy Faber, \& Ari Maller for useful
discussions.  We 
thank 
Andrey Kravtsov \& Anatoly Klypin for 
running and providing access to 
the ART simulations.  
This work has been supported by a GAANN fellowship to R.H.W., and by 
NSF \& NASA grants at UCSC.}


\begin{references}

\begin{small}

\reference Adelberger, K. L., Steidel, C. C., Giavalisco, M., 
Dickinson, M., Pettini, M., \& Kellogg, M. 1998, \apj, 505, 18 (A98)
%
\reference Baugh, C. M., Cole, S., Frenk, C. S., Lacy, C. G. 1998, \apj, 498,
504

\reference Bullock, J. S., Kolatt, T. S., Sigad, Y.,Somerville R. S., 
Kravtsov, A. V., Klypin A. A., Primack, J. R., Dekel, A 1999, \mnras,
submitted
 
\reference Governato., F. Baugh, C. M., Frenk, C. S., Cole, S., Lacey, C. G., Quinn, T. \& Stadel, J.,1998, Nature, 392, 359


\reference Jing, Y. P. \& Suto, Y. 1998, \apjl, 494, L5
 
\reference Kauffmann, G., Colberg, J. M., Diafero, A., \& White, S. D. M., 
1999, \mnras, 303, 188

\reference Katz, N, Hernquist, L, \& Weinberg, D. H., 1999, \apj, 523, 463

\reference Kolatt, T. S., Bullock, J. S.,Sigad, Y.,Somerville R. S., 
Kravtsov, A. V.,  Klypin A. A., Primack, J. R., Dekel, A 1999, \apjl, 523, 109

\reference Kravstov, A. V., Klypin, A. A., \& Khokhlov, A. M.1997, ApJS,111,73

\reference Lowenthal, J. D., \etal 1997, \apj, 481,673

\reference Mo, H. J. \& White, S. D. M. 1996, MNRAS, 282, 347



\reference Sheth, R. K. \& Tormen, G. 1999, \mnras, 308, 119

\reference Somerville, R. S., Primack, J. R., \& Faber, S. M. 
1999, \mnras, in press (SPF) 

\reference Steidel, C. C., Adelberger, K. L., Dickinson, M., 
Giavalisco, M., Pettini, M., \& Kellogg, M. 1998, \apj, 492, 428 (S98)

\reference Steidel, C. C., Adelberger, K. L., 
Giavalisco, M., Dickinson, M., Pettini, M.1999, \apj, 457, 645


\reference Wechsler, R. H., Gross, M. A. K., Primack, J. R.,
Blumenthal, G. R. \& Dekel, A. 1998, \apj, 506, 19 (W98) 

\reference Wechsler, R. H., Somerville, R. S., Bullock, J. S., Kolatt, T. S.,
Primack, J. R., Blumenthal, G. R. \& Dekel, A. in preparation (W99)

\end{small}

\end{references}
\end{document}